# Analyse klassischer und quantenmechanischer Wahrscheinlichkeitsbegriffe

## Eine Zusammenschau

**Christian Hugo Hoffmann**

House of Lab Science AG, Garstligweg 8, 8634 Hombrechtikon, Schweiz

Technopark Zurich, Technoparkstrasse 1, 8005 Zürich, Schweiz

Ethik-Zentrum der Universität Zürich, Zollikerstrasse 117, 8008 Zürich, Schweiz

*Abstract*: Dieser Beitrag widmet sich der zentralen Frage, wie ein kohärentes Konzept der Wahrscheinlichkeit aussehen könnte, das sowohl der klassischen Wahrscheinlichkeitstheorie, die durch Kolmogorov axiomatisiert wurde, als auch der Quantentheorie gerecht würde. In einer Zeit, wo Quanten verstärkte und ausgeweitete Aufmerksamkeit erfahren – man denke etwa an die Fortschritte bei Quantencomputern oder die Versprechen, die sich an diese neue Technologie knüpfen (National Academies of Sciences: Engineering, and Medicine, 2019), – sollte einer adäquaten Interpretation der nicht minder bedeutsamen Wahrscheinlichkeit auch und gerade mit Blick auf die Quantentheorie besonderes Gewicht zukommen.

# 1. Einleitung

> I had always been convinced that the problem of the interpretation of
> the quantum theory was closely linked with the problem of the inter-
> pretation of probability theory in general […]. (Popper 1959, S. 27)

Die 20er und 30er Jahre des vergangenen Jh. waren von nicht überschätzbarer Bedeutung für die Naturwissenschaften und darüber hinaus. Auf der einen Seite bildete sich eine neue Physik, die neue Quantentheorie heraus,[1] die durch Heisenbergs Arbeit über eine quantentheoretische Umdeutung kinematischer und mechanischer Beziehungen begründet wurde. Auf der anderen Seite präsentierte Kolmogorov erstmals ein axiomatisches System, das eine solide Basis für die moderne Wahrscheinlichkeitstheorie bot. Dass und wie diese beiden Hochpunkte der Wissenschaftsgeschichte, „[which] opened new gates, not just for science, but for human thinking in general" (Dvurečenskij 2009, S. vii), im Hinblick auf eine gleich zu bestimmende Kernfrage eng zusammenhängen, ist eine Aufgabe der vorliegenden Arbeit.

Wahrscheinlichkeiten spielen eine zentrale Rolle in jener auf Heisenberg zurückgehenden Quantenmechanik, die heute meist als die erfolgreichste physikalische Theorie gepriesen wird. Zum einen ist (zumindest unter Fachleuten) die Ansicht weit verbreitet, dass die Quantentheorie intrinsisch statistisch sei, d.h. die Untersuchung der elementaren Teilchen und ihre Interaktionen irreduzibel von probabilistischer Art sei und nicht auf eine zugrundeliegende deterministische Theorie verwiesen werden könne. Auch wenn man zum anderen diese Auffassung nicht teilt,[2] bleiben Wahrscheinlichkeiten insofern zentral, dass der quantenmechanische Formalismus es uns gestattet, die Wahrscheinlichkeiten für mikroskopische Ereignisse zu berechnen. Darum scheint für die Förderung des Verständnisses der Quantentheorie – was gleichzeitig eine immense Herausforderung darstellt[3] – die Frage danach, was Wahrscheinlichkeit / mathematische Wahrscheinlichkeitsausdrücke überhaupt und im Rahmen der Quantenmechanik bedeuten, eine essentielle zu sein.

---

[1] Das Geburtsjahr der Quantentheorie überhaupt ist auf Plancks Entdeckung des berühmten Strahlungsgesetzes für den schwarzen Körper im Jahre 1900 datiert. Hierbei und bei Einsteins Erklärung des photoelektrischen Effekts spricht man – in Abgrenzung zu den seit 1925 stattfindenden Entwicklungen (zu dieser Zeit veröffentlichte Heisenberg sein Werk) – von der *alten* Quantentheorie. Wir nehmen mit dieser Unterscheidung (alt vs. neu) also nicht Bezug auf verschiedene Interpretationsvarianten der Quantenmechanik, wie z.B. die Kopenhagener Deutung. Fortan wird mit „Quantentheorie", „-mechanik" bloß die Rede von der neuen Quantentheorie sein.

[2] Es existieren daneben deterministische Interpretationen der Quantenmechanik, z.B. die De-Broglie-Bohm-Theorie, welche seit den 1990ern einen neuen Aufwind erlebt. Vgl. dazu etwa Holland (1993).

[3] Es sei an Feynmans Ausspruch erinnert: „[…] I think I can safely say that nobody understands quantum mechanics", vgl. Feynman (1965).

Die *Kernfrage* lautet in folgender Formulierung: Wie könnte ein kohärentes Konzept der Wahrscheinlichkeit aussehen, das sowohl der klassischen Wahrscheinlichkeitstheorie, die durch Kolmogorov axiomatisiert wurde,[4] als auch der Quantentheorie gerecht würde? Die Kernfrage beleuchtet somit den Zusammenhang zwischen der Quantentheorie und den Grundlagen der Wahrscheinlichkeit. Es soll um eine adäquate Interpretation der Wahrscheinlichkeit gehen bzw. um eine Analyse verschiedener Wahrscheinlichkeitsbegriffe auch und gerade mit Blick auf die Quantentheorie. Man sollte sich dabei bewusst machen, dass das Interpretieren der Wahrscheinlichkeit beinhaltet, das oder ein formales System der Wahrscheinlichkeit zu verstehen (z.B. indem man den Ausdrücken in den Axiomen und Theoremen vertraute Bedeutungen zufügt); da es aber nicht *das* formale System der Wahrscheinlichkeit gibt (sondern eine Schar solcher Systeme), schließt sich die Frage an, von welchem wir aus-gehen sollen.

Einerseits liegt es nahe, den Kolmogorovschen Wahrscheinlichkeitskalkül heranzuziehen, der nicht nur unter Philosophen den Status als Standard innehat; andererseits müsste dann geklärt werden, ob und inwiefern dieser mit der Quantentheorie vereinbar wäre.

Diese Arbeit entzieht sich diesen schwierigen Fragen, worin gleichzeitig eine ihrer Stärken liegt. Sie erhebt den Anspruch, für beide, sich gegenseitig ausschließende Fälle – also sowohl für konsistente *mögliche Welten*[5], in denen Kolmogorovs Axiome erfüllt sind, als auch für diejenigen, in welchen sie nicht erfüllt sind,[6] – zu zeigen, dass die prominentesten Interpretationen der Wahrscheinlichkeit unbefriedigend und problematisch sind.

Bevor dieses Fazit aus der im Teil I zu subjektiven Wahrscheinlichkeitsbegriffen und der im Teil II zu objektiven Wahrscheinlichkeitsbegriffen durchzuführenden Analyse gezogen wird, sollen zunächst, in Abschnitt 2, einige Kriterien zur Sicherstellung der Adäquatheit der Interpretationen der Wahrscheinlichkeit eingeführt werden, wobei wir uns v.a. an den Ausführungen von Salmon (1967) orientieren wollen.

---

[4] Freilich gibt es neben der Kolmogorovschen Axiomatisierung der Wahrscheinlichkeitstheorie alternative axiomatische Konstruktionen. Jedoch denken Philosophen typischerweise an Kolmogorov (1933), wenn sie von *Wahrscheinlichkeitstheorie* sprechen. Vgl. Hájek (2001).
[5] Vgl. für eine ausführliche Darstellung des Grundgedankens der *Möglichen-Welten-* oder *Ähnlichkeitssemantik* Lewis (1973), Abschn. 1.1 – 1.4.
[6] Warum diese Unterscheidung vorgenommen wird, soll in Abschnitt 2 ausführlicher erläutert werden. Ebenso folgen dort einige Anmerkungen zu Kolmogorov (1933).

## 2. Kriterien für die Adäquatheit der Interpretationen der Probabilität

Welche Kriterien sind geeignet, um die Stichhaltigkeit einer vorgeschlagenen Interpretation der Wahrscheinlichkeit zu beurteilen? Außer Frage steht, dass eine Interpretation nicht-zirkulär, widerspruchsfrei, präzise, eindeutig sein sollte und (womöglich) einfache, gut verständliche Grundbegriffe gebraucht. Dies ist kennzeichnend für gutes Philosophieren im Allgemeinen. Welche speziellen Anforderungen sollten jedoch vernünftigerweise an Interpretationen der Wahrscheinlichkeit gestellt werden? Salmon (1967, S. 63f.) schreibt dazu, was wir im Folgenden kommentieren und ggf. ergänzen wollen:

> [1] *Admissibility.* We say that an interpretation of a formal system is admissible if the meanings assigned to the primitive terms in the interpretation transform the formal axioms, and consequently all the theorems, into true statements. A fundamental requirement for probability concepts is to satisfy the mathematical relations specified by the calculus of probability…
>
> [2] *Ascertainability.* This criterion requires that there be some method by which, in principle at least, we can ascertain values of probabilities. It merely expresses the fact that a concept of probability will be useless if it is impossible in principle to find out what the probabilities are…
>
> [3] *Applicability.* The force of this criterion is best expressed in Bishop Butler's famous aphorism, "Probability is the very guide of life."…

Gemäß Salmon (ebd. S. 63) ist zu konstatieren: „[If the] three criteria [are] fulfilled, […] we are to have a satisfactory interpretation of probability."

Ad 1: Auf den ersten Blick scheint dieses Kriterium an einen Allgemeinplatz anzuknüpfen: Erstens würden ‚Interpretationen' eines Wahrscheinlichkeitskalküls, die z.B. das Wahrscheinlichkeitsmaß p als „Anzahl der Schüler in der Schule" deuten, die entsprechenden Axiome und Theoreme offenkundig falsch wiedergeben. Und zweitens wird das Wort „Interpretation" meist auf eine Weise verwendet, dass „admissible interpretation" einen Pleonasmus darstellt.

Auf den zweiten Blick ist das Kriterium jedoch keinesfalls trivial: Erstens existieren neben der klassischen Kolmogorovschen Axiomatisierung (wie angedeutet) zahlreiche andere und es herrscht kein Konsens darüber, welche die ‚beste' ist (was auch immer das heißen mag).[7] Und zweitens bleibt bei bestimmten führenden Interpretationen der Wahrscheinlichkeit – das gilt gerade für manche, für die Quantentheorie scheinbar so wichtigen Propensität-Interpretationen – mindestens der Status einiger seiner Axiome unklar. In nuce: Es gibt nicht die *admissibility* tout court, sondern vielmehr besteht sie (oder eben nicht) im Hinblick auf dieses oder jenes Axiomensystem.

Dieses erste Kriterium sollte allerdings nicht übergewichtet werden. Im Gegenteil, die Tatsache, dass mehrere vorgebrachte Axiomatisierungen Inkonsistenzen enthalten – Carnap (1950, S. 341) weist dies bspw. Jeffreys (1939) nach – und die Diskussion der Vorzüge und Schwächen der daran anschließenden Interpretationen der Wahrscheinlichkeit gleichwohl wertvoll und fruchtbar sein kann, sollte schon klar machen, dass wir gut daran tun, unseren Fokus auf andere Kriterien zu legen. Darum spielt es auch keine erhebliche Rolle, wenn wir einfach undifferenziert mögliche Welten, in denen Kolmogorovs Axiome nicht gelten, möglichen Welten gegenüberstellen, in denen sie gelten, um die in Kürze vorzustellenden Wahrscheinlichkeitsbegriffe für alle Fälle bewerten zu können.[8]

Ad 2: Abgesehen davon, dass ein wenig unklar ist, worauf Salmon (1967) mit seiner Wendung „in principle" hinauswill,[9] ist die Aussage des Kriteriums zweifellos unstrittig.

Ad 3: Auf den Stellenwert seines dritten Kriteriums macht Salmon selbst aufmerksam (ebd. S. 65): „An explication that fails to fulfill the criterion of applicability is simply not an explication of the concept we are trying to explicate." Die Konzepte von Länge, Volumen oder Flächeninhalt z.B. tragen allesamt [1] und [2] Rechnung – sie fallen nämlich unter die mathematische Maßtheorie –; gleichzeitig gelten sie sicherlich als nützliche Begriffe, sie können insofern aber als *guides of life* angesehen werden. Ergo würden sie als Interpretationen der Wahrscheinlichkeit im strikten Sinne taugen (weil es eben den Anschein

---

[7] In der Tat weicht sogar Salmons präferiertes Axiomensystem von Kolmogorovs ab (vgl. ebd. S. 59). Letzteres lässt sich wie folgt umreißen: Sei $\Sigma$ eine nicht-leere, endliche Menge (*Ergebnisraum*). Eine Funktion p, die den Teilmengen von $\Sigma$ (*Ereignisalgebra* oder *Ereignisraum*) reelle Zahlen zuordnet und die Eigenschaften

| (K1) | $p(A) \geq 0$, | (Nicht-Negativität) |
|---|---|---|
| (K2) | $p(\Sigma) = 1$, | (Normierungsaxiom) |
| (K3) | für beliebige disjunkte Mengen A, B $\subset \Sigma$ gilt: | |
| | $p(A \cup B) = p(A) + p(B)$ | ((Finites) Additionsaxiom) |

hat, heißt Wahrscheinlichkeitsmaß auf der Potenzmenge von $\Sigma$ und die Funktionswerte von p heißen Wahrscheinlichkeiten. Diese aufgeführten Regeln (K1 – K3) entsprechen den Axiomen von Kolmogorov.
Alternative Axiomatisierungen geben etwa K2, K3 (oder die Verallgemeinerung K3', die auch unendliche Räume erfasst) auf, führen Wahrscheinlichkeiten ein, die infinitesimale Werte annehmen oder die vage (z.B. übersetzt durch Mengen numerischer Werte) sein können. Vgl. weiterführend Rényi (1970), Fine (1973), Skyrms (1980).
[8] Auf eine auf die Quantenmechanik bezugnehmende Motivation für die getroffene Zweiteilung wird in Abschnitt 3 noch eingegangen.
[9] Vielleicht soll sie einfach nur zur Ausschließung von Allwissenheit dienen.

hat, dass alle drei Kriterien erfüllt sind). Um diesen Schluss zu vermeiden, um also ungewünschte ‚Interpretationen' der Wahrscheinlichkeit auszuschließen, wollen wir [3] etwas näher betrachten (was auch Salmon tut), dabei Hájek (2007b) folgen und mehrere Folgekriterien unter [3] subsumieren.[10]

Nicht-Trivialität: Eine Interpretation sollte nicht-extreme Wahrscheinlichkeiten (die also ungleich null und ungleich eins sind) berücksichtigen.[11]

Anwendbarkeit auf Häufigkeiten: Eine Interpretation sollte die Beziehung zwischen Wahrscheinlichkeiten und Häufigkeiten (*in the long run*) verdeutlichen.[12]

Anwendbarkeit auf rationale Glaubensgrade: Eine Interpretation sollte die Rolle klar herausstellen, welche die Wahrscheinlichkeiten in der Einschränkung von Glaubensgraden spielen.[13]

Anwendbarkeit auf die Wissenschaften: Eine Interpretation sollte die paradigmatischen Gebräuche der Wahrscheinlichkeit in den Wissenschaften (v.a. in der Quantenmechanik) beleuchten.

Anwendbarkeit auf Erweiterungsschlüsse: Eine Interpretation kann darüber hinaus damit bestechen, dass sie die Unterscheidung zwischen *guten*[14] und *schlechten*[15] Erweiterungsschlüssen klarlegt und die Gegenüberstellung zu deduktiven Schlüssen expliziert.

Vielleicht mag man den Wahrscheinlichkeitsbegriffen weitere metaphysische Bedingungen auferlegen; bspw. scheint ein Zusammenhang zwischen Wahrscheinlichkeit und Modalität zu bestehen: Ereignisse mit positiver Wahrscheinlichkeit können stattfinden, selbst wenn sie tatsächlich nicht geschehen.[16] Nichtsdestotrotz reicht der – in Anlehnung an Salmon (1967) und Hájek (2007b) – erstelle Katalog an Kriterien und Subkriterien aus, um im Folgenden die führenden Interpretationen gewinnbringend untersuchen und bewerten zu können.

---

[10] Hájek möchte sie nicht als notwendige Bedingungen verstanden wissen. Ob sie zusammengenommen (inklusive [1] und [2]) hinreichend für eine (gelungene) Interpretation der Wahrscheinlichkeit sind, würde er vermutlich bejahen.

[11] Zur Illustration liefert Hájek (ebd.) nachstehendes Bsp.: „[…] suppose that we interpret '*p*' as the *truth* function: it assigns the value 1 to all true sentences, and 0 to all false sentences. […] all the axioms come out true […]. We would hardly count it as an adequate *interpretation of probability*, however, and so we need to exclude it."

[12] Es geht insbesondere um die Beantwortung der Frage, warum wahrscheinlichere Ereignisse öfters geschehen als weniger wahrscheinliche?

[13] U.a. insofern, dass ein rationaler *Agent* stärker an das Eintreten des wahrscheinlicheren Ereignisses glaubt.

[14] Etwa derart: Die Zuckerstücke $Z_1$, …, $Z_n$ besitzen eine Disposition, sich in Wasser aufzulösen; also besitzt auch $Z_{n+1}$ diese Disposition.

[15] Etwa derart: Ich besuche ein Seminar zur Stochastik, also existiert das fliegende Spaghettimonster.

[16] Interessant und kontrovers ist, ob auch die Umkehrung gilt: Nur Ereignisse mit positiver Wahrscheinlichkeit können geschehen. Vgl. weiterführend dazu Hájek (2011).

### 3. Teil I: Führende Interpretationen klassischer oder quantenmechanischer Wahrscheinlichkeiten

> The idea of probability leads in two different directions.
> (Hacking 2008, S. 127)

Grundsätzlich macht es das klare Nachdenken über Wahrscheinlichkeit erforderlich, zwei Grundströmungen auseinander zu halten: Auf der einen Seite vereinen sich Interpretationen unter dem Titel der subjektiven Wahrscheinlichkeitsbegriffe und auf der anderen Seite steht eine Klasse objektiver Wahrscheinlichkeitsbegriffe. Diese Grundunterscheidung möchten wir hier aufgreifen, indem wir bezüglich der Kernfrage eine zweiteilige Analyse durchführen (Abschnitt 3 und 4). In diesem ersten Teil wenden wir uns subjektiven Wahrscheinlichkeitsbegriffen zu, welche Glaubensgrade hervorheben (ebd. S. 140). Was ist darunter zu verstehen? Popper (1959, S. 25) gibt darauf die befriedigende Antwort – und bringt damit gleichzeitig vor, was diesen Interpretationen gemeinsam ist: „Probability theory is regarded as a means of dealing with the *incompleteness of our knowledge*, and the number [*p(a)*, i.e. the probability of the event *a*, C.H.] is regarded as a measure of the degree of rational assurance, or of rational belief, […].“[17]

Eine zweite, für die Analyse *speziell* wichtige Unterscheidung wurde zuvor schon getroffen und soll an dieser Stelle nochmals, aber diesmal aus anderer Sicht plausibilisiert werden. Noch nicht hinreichend klar ist nämlich, warum (i) in der Analyse überhaupt zwischen der durch Kolmogorov axiomatisierten klassischen Wahrscheinlichkeitstheorie und allen anderen Theorien der Wahrscheinlichkeit unterschieden wird, für welche diese Postulate nicht gelten, und warum (ii) nicht statt letzterem eine geeignete Theorie oder geeignete Theorien der *Quantum probability* genau benannt respektive präsentiert werden.[18] Scheint es sogar nicht (iii) gemäß Salmons *Admissibility*-Kriterium unumgehbar, aufzuwerfen, was den Wahrscheinlichkeitskalkül der Quantenmechanik auszeichnet, wenn sich der *Kernfrage* angenommen werden, und damit erforscht werden soll, ob ein sowohl für die klassische

---

[17] Popper spricht, statt von *p(a)*, von dem Maß *p(a, b)*, was der konditionalen Wahrscheinlichkeit entspricht. Vgl. tiefergehend im Felde der personalen Wahrscheinlichkeit: de Finetti (1980), Kemeny (1955), Ramsey (1926) (wobei die Beiträge folgende Schlüsselkonzepte besprechen: Wett-Interpretationen, Dutch-Book-Argumente, Bayesianismus) und im Felde der logischen Wahrscheinlichkeit: Keynes (1921) und v.a. Carnap (1950).
[18] Wertvolle Beiträge zur *Quantum probability* stammen etwa von Gudder (2009) oder von Wilce (2010).

Wahrscheinlichkeitstheorie als auch für die Quantentheorie passender (d.h. insbesondere den aufgestellten Kriterien genügender) Wahrscheinlichkeitsbegriff existiert?

Zwar wurde schon oben dafür argumentiert, Salmons erstes Kriterium nicht zu streng auszulegen, doch kann auch auf andere Weise (ii) und (iii) der Nährboden entzogen werden und gleichzeitig mit einer Erwiderung auf (i) die getroffene Zweiteilung (gänzlich) legitimiert werden: Die Quantenmechanik veranlasst zur Aufgabe der meist als Standard anerkannten Kolmogorovschen Axiome (ad i), was zu einer Fallunterscheidung im Sinne der hier propagierten anregt; sie zwingt aber nicht dazu (ad ii und iii), sodass man zum Zwecke der Einfachheit und Übersichtlichkeit und selbst bei einer ‚strengen Lesart' des *Admissibility*-Kriteriums keine alternativen Fälle gegenüber den zwei vorgeschlagenen Fällen zu betrachten braucht.

> One can argue that there is nothing wrong with our traditional concept of probability. The apparent violation of its rules [by quantum mechanics or, more precisely, by quantum frequencies, C.H.] is caused by »measurement disturbances«. Now, quantum theory does not speak about measurement disturbances but rather about »interference«, and the meaning of this term is precisely what is at stake. Quantum mechanics does not provide any dynamic *mechanism* of measurement disturbances. If one introduces such a mechanism one goes a step beyond quantum theory and obtains a non local theory.[19] As yet there is no independent evidence for the existence of such a mechanism, the only evidence is the violation of Bell inequalities[20] themselves. Thus the dilemma is to choose between non local theories and the axioms of probability. Presently no rational guideline for such a choice seems to be available. (Pitowsky 1989, S. 183)

Es ist also – quasi als gesicherter Minimalkonsens – festzuhalten, dass die üblichen Axiome der Wahrscheinlichkeitstheorie nicht als *analytisch* gelten können, mit anderen Worten: „[…] the axioms do not necessarily capture the proper meaning of the term »probability« (though they surely represent a possible analysis of the term)" (ebd. S. 183f.). Die teils verheerenden Folgen aus dieser Einsicht für bestimmte Schulen wollen wir im Fortgang und unter Zuhilfenahme der Fallunterscheidung aufzeigen.

---

[19] Das *Lokalitätsprinzip* besagt – vereinfacht gesprochen – dass die Messung einer Eigenschaft an einem Teilchen nicht die Messung an einem anderen entfernten Teilchen beeinflussen kann.
[20] Für Hintergründe zu den sogenannten Bellschen Ungleichungen vgl. Bell (1964) und z.B. Shimony (2009).

### 3.1.    Personale Wahrscheinlichkeiten

Sicherlich mag die mit *belief dogmatists*[21] etikettierte Schule der Subjektivisten eine der erfolgreichsten sein. Deshalb ist es auch nicht weiter verwunderlich, dass die diese Beobachtung würdigende ‚Anwendbarkeit auf rationale Glaubensgrade' – in Übereinstimmung mit Hájek (2007b) – als Kriterium in Abschnitt 2 erhoben wurde. Nichtsdestotrotz scheint diese Klasse subjektiver Interpretationen der Wahrscheinlichkeit mehr in Bezug auf die klassische Wahrscheinlichkeitstheorie, weniger jedoch hinsichtlich der Quantentheorie zu überzeugen, wobei wir uns in diesem Urteil auf von Neumann (1943) stützen und sie darum in dieser Arbeit einfachheitshalber (auch um ihren Rahmen nicht zu sprengen) nicht weiter berücksichtigen wollen; damit sei aber wohlbemerkt nicht ausgeschlossen, dass sie für die Quantentheorie (oder vice versa) fruchtbar gemacht werden könnte.[22]

Bereits in seinem Vorwort (zu 1943) antizipiert von Neumann das Programm und die Konklusion, die sich für die Möglichkeit der ‚Neutralisierung' des statistischen Charakters der Quantenmechanik ergebe und sondert insofern strenge subjektive Wahrscheinlichkeitsbegriffe aus:

> There will be a detailed discussion of the problem as to whether it is possible to trace the statistical character of quantum mechanics to an ambiguity (i.e., incompleteness) in our description of nature. Indeed, such an interpretation would be a natural concomitant of the general principle that each probability statement arises from the incompleteness of our knowledge.
> This explanation "by hidden parameters" [...] has been proposed more than once. However, it will appear that this can scarcely succeed in a satisfactory way, or more precisely, such an explanation is incompatible with certain qualitative fundamental postulates of quantum mechanics.

---

[21] „This leads to the idea of personal probability. It is a kind of belief-type probability, [which] is totally 'subjective'" (Hacking 2008, S. 142), was etwa in Aussagen folgender Art zum Ausdruck kommt: „Ich persönlich bin sehr davon überzeugt, dass Bob der kleinen Mimi die Schokolade weggenommen hat"; oder „Wenn ich darauf wetten müsste, würde ich 9 zu 1 darauf setzen, dass Bob der kleinen Mimi die Schokolade weggenommen hat." Vgl. für diese Form der Klassifizierung ebd. S. 140ff.

[22] Viele derer, die etwa für eine Einführung von *hidden variables* in die Quantentheorie plädieren (z.B. Albert Einstein, in diesem Zusammenhang kann auch sein berühmtes Diktum: „Gott würfelt nicht" gesehen werden, vgl. auch Einstein/Podolsky/Rosen (1935)), sprechen sich auch für einen subjektiven Wahrscheinlichkeitsbegriff aus. Darum ist es wichtig zu sehen, dass in dem Sinne eine ‚starke Annahme' gemacht wurde, dass bestimmte Interpretationen der Quantenmechanik (wohl aber nicht die gängigsten) durchaus auf strenge subjektive Wahrscheinlichkeitsbegriffe rekurrieren (können). Diese Tatsache soll keinesfalls verschwiegen, sie kann nur aufgrund der Möglichkeiten dieser Arbeit nicht ausgewalzt werden.
Für einen Überblick über die Interpretation der Wahrscheinlichkeit als personale Wahrscheinlichkeit relativ zur klassischen Wahrscheinlichkeitstheorie, der auch kritische Aspekte nicht ausspart, vgl. v.a. Ramsey (1926).

## 3.2.   Wahrscheinlichkeitslogischer Ansatz, von Neumann II. (1937/1954)

Wahrscheinlichkeitslogische Ansätze, die berühmte Proponenten vorweisen können (in erster Linie Keynes 1921 und Carnap 1950), gehören zwar nach unserer (von Hacking 2008 übernommenen) Klassifizierung ebenfalls in das Feld der Subjektivisten,[23] gleichwohl scheinen sie – selbst in von Neumanns Augen – merklich interessanter und geeigneter für die Quantentheorie zu sein. Und mehr noch, er liebäugelt gar mit einer *logical theory of probability*, die er mit J. M. Keynes (1921) verbindet:[24]

> Essentially if a state of a system is given by one vector, the transition probability in another state is the inner product of the two which is the square of the cosine of the angle between them. In other words, probability corresponds precisely to introducing the angles geometrically. Furthermore, there is only one way to introduce it. The more so because in the quantum mechanical machinery the negation of a statement, so the negation of a statement which is represented by a linear set of vectors, corresponds to the orthogonal complement of this linear space.
>
> And therefore, as soon as you have introduced into the projective geometry the ordinary machinery of logics, you must have introduced the concept of orthogonality. ... In order to have probability all you need is a concept of all angles, I mean angles other than 90°. Now it is perfectly quite true that in geometry, as soon as you can define the right angle, you can define all angles. Another way to put it is that if you take the case of an orthogonal space, those mappings of this space on itself, which leave orthogonality intact, leave all angles intact, in other words, in those systems which can be used as models of the logical background for quantum theory, it is true that as soon as all the ordinary concepts of logic are fixed under some isomorphic transformation, all of probability theory is already fixed.
>
> What I now say is not more profound than saying that the concept of a priori probability in quantum mechanics is uniquely given from the start. ... This means, however, that one has a formal mechanism, in which logics and probability theory arise simultaneously and are derived simultaneously.
> (von Neumann 2001, S. 244f.)

Von Neumann war fasziniert von der Bestimmtheit der Wahrscheinlichkeit durch die Logik im Kontext der Quantenmechanik, aber er erachtete diese *logical theory* (Interpretation) der

---

[23] Andere Autoren (z.B. Pitowsky 1989) bilden neben der Oberklasse der Subjektivisten respektive der Objektivisten eine dritte Oberklasse für die Anhänger der logischen Wahrscheinlichkeit. Für beide Verfahren lassen sich gute Gründe finden. Carnap selbst sah irrtümlicherweise die logische Wahrscheinlichkeit als objektiv an.

[24] Seine ausbuchstabierten Gedanken dazu wurden leider erst und lediglich beim Internationalen Mathematikerkongress 1954 in Amsterdam übermittelt. In einem unfertigen Manuskript „Quantum logic (strict and probability logics)" (geschrieben um 1937) macht von Neumann allerdings bereits entsprechende Andeutungen.

Wahrscheinlichkeit als noch nicht vollständig ausgereift: er erwähnt die Notwendigkeit einer Axiomatisierung als eines der offenen Probleme der Mathematik (Amsterdam 1954, vgl. ebd.).

Die bei weitem umfangreichste und systematischste Abhandlung über logische Wahrscheinlichkeit (allerdings nicht vor dem Hintergrund der Quantenmechanik) stammt von Carnap (1950). Seine Formulierung der logischen Wahrscheinlichkeit beginnt mit der Konstruktion einer formalen Sprache.[25] Er definiert dann Zustandsbeschreibungen (*state descriptions*) als die stärksten konsistenten Aussagen (in einer gegebenen Sprache), welche alle Individuen – so detailliert wie es dem Ausdrucksvermögen der Sprache möglich ist – beschreiben.[26] Bezeichnen wir weiter als Strukturbeschreibung (*structure description*) eine maximale Menge von Zustandsbeschreibungen, wobei jede aus einer anderen durch eine bestimmte Permutation von individuellen Namen hervorgehen kann.[27] Ein beliebiges Wahrscheinlichkeitsmaß p (–) über die Zustandsbeschreibungen weitet sich automatisch zu einem Maß über alle Sätze aus, da jeder Satz H äquivalent zu einer Disjunktion von Zustandsbeschreibungen ist, wodurch seine (*Prior-)Wahrscheinlichkeit* p (H) determiniert ist; p wiederum induziert eine Bestätigungsfunktion (*confirmation function*) c (–, –), für die Carnap (1963) ferner einige Axiome festschreibt:[28]

$$c\,(H, E) = \frac{p\,(H \wedge E)}{p\,(E)}.$$

Anders ausgedrückt: Wir ermitteln den als (induktive) Wahrscheinlichkeit ausgedrückten Stützungsgrad zwischen einer Evidenz E und einer gegebenen Hypothese H:[29] „The asserted probability value means the degree to which the hypothesis is confirmed or supported by the evidence" (Carnap 1953, S. 192). Schließlich bringt p einen Grad der Bestätigung c hervor, der *Lernen aus der Erfahrung* ermöglicht und erfasst. Wahrscheinlichkeitslogische Aussagen dieser Art, also von der Art „c (H, E) = x", hielt Carnap für Tautologien (*logical truths*).

---

[25] Genauer: Er erwägt eine Klasse sehr simpler Sprachen, die aus einer endlichen Zahl an logisch unabhängigen einstelligen Prädikaten (*naming properties*) besteht, angewandt auf abzählbar viele Individuenkonstanten (*naming individuals*) oder Variablen und die üblichen Junktoren.

[26] Bspw. sei eine Sprache gegeben mit zwei Namen a und b für Individuen, dem Prädikat $F^1$ und folgenden Zustandsbeschreibungen: 1) Fa ∧ Fb; 2) ¬Fa ∧ Fb; 3) Fa ∧ ¬Fb; 4) ¬Fa ∧ ¬Fb.

[27] In unserem Bsp. (aus der vorherigen Fußnote) gibt es drei Strukturbeschreibungen: {1}, „Alles ist F"; {2, 3}, „ein F, ein ¬F"; und {4}, „Alles ist ¬F".

[28] Darunter finden sich diejenigen, die für den Kolmogorovschen Wahrscheinlichkeitskalkül selbst gelten, und verschiedene Axiome der Symmetrie (z.B., dass c (H, E) unter Permutationen von Individuen unverändert bleibt).

[29] Offenkundig existieren unendlich viele Kandidaten für p und somit auch für c (selbst für sehr simple Sprachen). In (1950) argumentiert Carnap für sein favorisiertes Maß p\*, das jeder Strukturbeschreibung das gleiche Gewicht zuschreibt, welches wiederum in gleichem Maße unter den die Strukturbeschreibungen konstituierenden Zustandsbeschreibungen aufgeteilt wird. Demnach werden unsere drei Strukturbeschreibungen jeweils mit 1/3 gewichtet und p\* lässt sich für die einzelnen Zustandsbeschreibungen wie folgt bestimmen: p\* (Fa ∧ Fb) = 1/3 × 1 = 1/3; p\* (¬Fa ∧ Fb) = 1/3 × 1/2 = 1/6; p\* (Fa ∧ ¬Fb) = 1/3 × 1/2 = 1/6; und p\* (¬Fa ∧ ¬Fb) = 1/3 × 1 = 1/3.
Betrachten wir nun die Hypothese H: Fb, die in zwei von vier Zustandsbeschreibungen wahr ist, mit der A-priori-Wahrscheinlichkeit p\* (H) = 1/2. Angenommen wir überprüfen das Individuum *a* und finden heraus, es hat die Eigenschaft F – dies sei Evidenz E mit p\* (E) = 1/2. E ist ein schwacher positiver *induktiver* Beleg für H. Es gelte nunmehr: p\* (H ∧ E) = 1/3 (weil p\* (Fb ∧ Fa) = 1/3), sodass für die Bestätigungsfunktion c\* (H, E) folgt:
$$c\* (H, E) = p\* (H \wedge E)\,/\,p\* (E) = 2/3.$$
Dieser Wert (a posteriori) entspricht der bedingten Wahrscheinlichkeit von H, gegeben E und ist größer als die A-priori-Wahrscheinlichkeit p\* (H) = 1/2; ergo wurde die Hypothese bestätigt/gestützt.

Ohne Einzelheiten seiner Arbeiten darzulegen kann dank des Anreißens der wesentlichen Punkte seiner Konzeption, das die sonst schwerer greifbaren wahrscheinlichkeitslogischen Theorien veranschaulichen sollte, für die logischen Interpretationen insgesamt zusammengefasst werden: (i) Ihnen geht es darum, das Stützungsverhältnis oder den Grad der Bestätigung von einer Evidenz E und einer gegebenen Hypothese H in voller Allgemeinheit einzufangen. (ii) Insofern kann auch eine erhellende Rede vom *Grad der Implikation* sein, der E mit H verknüpft und der Carnaps Programm der *induktiven Logik* ausmacht.[30] Und (iii) werden sie da-durch herausgestellt, dass Wahrscheinlichkeiten a priori bestimmt werden können, indem man den *Ergebnisraum* prüft (oder um es mit Carnap zu halten, indem man sämtliche Zustandsbeschreibungen betrachtet und sie zueinander ins Verhältnis setzt).[31]

Nun soll die Carnapsche Theorie und mit ihr das ganze Lager seiner Anhänger kritisch in Au-genschein genommen werden. In der Literatur wurden viele gute Einwände in die Diskussion um die Überzeugungskraft des wahrscheinlichkeitslogischen Ansatzes eingebracht (vgl. dazu etwa Scott / Krauss 1966, Earman 1992 und Hájek 2001). Jedoch sind die Folgen aus unserer (dank Pitowsky 1989 gewonnenen) Einsicht, dass die Kolmogorovschen Axiome der Wahrscheinlichkeit nicht mit Notwendigkeit gelten (siehe oben), desaströs für Carnap und Konsorten:

> Carnap (1950) took the axioms of probability as mere conclusions of propo-sitional logic. As such, these axioms should be valid in all possible worlds. Since they are not, then either one has to give up the logical conception of probability, or else to deny the analytic character of logic. Both alternatives undermine Carnap's position. (Pitowsky 1989, S. 184)

---

[30] Der Name „induktive Logik" mag etwas unglücklich gewählt sein, nicht weil E nicht eine (enumerativ) induktive Evidenz sein muss – was gerne vorgebracht wird, was aber de facto nur zeigt, dass Carnap ein weiteres Verständnis des Induktionsbegriffs hat als diese Kritiker –, sondern weil er eine Schwarzweißzeichnung zwischen der deduktiven Logik einerseits mit ihrem (klassischen) Implikationsbegriff und der induktiven Logik anderer-seits mit ihrem Begriff des Implikationsgrades heraufbeschwört, die in die Irre führt. Denn die Fälle, in denen ein klassisches Implikationsverhältnis besteht bzw. nicht besteht, spiegeln nichts weiter als die Extremfälle wider, wo die Bestätigungsfunktion die Werte 1 bzw. 0 annimmt. Carnap wäre vermutlich nicht empfänglich für diese Kritik, weil er mit guten Gründen den gewählten Namen verteidigen könnte (vgl. Carnap 1953, S. 189f.).
Ungeachtet dieses Streits um Worte ist vielmehr der springende Punkt, dass die logische Interpretation der Wahrscheinlichkeit eine Rahmenordnung für Induktion bereitstellt, einen Weg „to explicate common ways of inductive reasoning […]" (ebd. S. 190). Ob Carnap mit seiner induktiven Logik die Induktionsprobleme (insbesondere das Humesche Induktionsproblem) lösen könnte, steht (wortwörtlich) auf einem ganz anderen Blatt.
[31] In Abgrenzung zu den frühen Wahrscheinlichkeitstheoretikern wie Laplace müssen die Möglichkeiten (oder die möglichen Ergebnisse eines Zufallsexperiments) nicht gleichgewichtet werden.

Es reicht also für die logische Wahrscheinlichkeitsauffassung im Sinne Carnaps nicht aus, dass Kolmogorovs Axiome in der aktualen Welt oder in einigen möglichen Welten logisch wahre Aussagen bilden, sondern sie müssten notwendig wahr sein, was wir gerade unter Berufung auf Pitowsky bestreiten. Infolgedessen ist zum einen unsere Fallunterscheidung zwischen Welten, in denen die klassischen Axiome für Wahrscheinlichkeiten gültig, und denen, wo sie ungültig sind, hier überflüssig und wird deshalb nicht durchgeführt; und zum anderen scheint gar der wahrscheinlichkeitslogische Ansatz im Lichte der Erkenntnisse hinfällig zu sein – ein Dilemma bliebe für Carnap zurück –, weshalb wir uns nicht weiter mit ihm aufhalten werden.

## 4. Teil II: Führende Interpretationen klassischer oder quantenmechanischer Wahrscheinlichkeiten

Nachdem unsere Suche nach einem kohärenten Konzept der Wahrscheinlichkeit für sowohl die klassische Wahrscheinlichkeitstheorie als auch für die Quantentheorie bisher im Revier der Subjektivisten recht erfolglos verlief – selbst der für von Neumann (1937/1954) vielversprechende wahrscheinlichkeitslogische Ansatz stellte sich zumindest in seiner am besten ausgearbeiteten Form als unzulänglich heraus –, widmen wir uns jetzt in diesem zweiten Teil der Analyse den Feldern der *objektiven* Wahrscheinlichkeit. Objektive Wahrscheinlichkeitsbegriffe geben *prima facie* geeignetere Kandidaten für zumindest die Quantenmechanik ab. Dies mag vielleicht daran liegen, dass sie im Gegensatz zu den subjektiven Konzepten Wahrscheinlichkeiten ‚in der Welt' lokalisieren, was gut mit dem scheinbar statistischen Charakter der Quantentheorie vereinbart werden kann. Erneut lassen sich die objektiven Auffassungen in den Worten Poppers (1959, S. 25) durch folgendes Merkmal beschreiben: „they all interpret [*p(a)* = r, C.H.][32] as a statement that can, in principle, be objectively *tested*, by means of statistical tests."

Eine in unseren Maßstäben gelungene Interpretation der Wahrscheinlichkeit soll aber (und neben ihrem mutmaßlichen *initial appeal*) der Quantentheorie *wirklich*, auch der klassischen Wahrscheinlichkeitstheorie und v.a. unseren, in Abschnitt 2, aufgestellten Kriterien Genüge tun. Wir werden im Fortgang sehen, dass dieses Desiderat auch für die prominenten objektiven Wahrscheinlichkeitsbegriffe nicht hinreichend erfüllt ist und sie folglich revisionsbedürftig sind.

### 4.1. Relative Häufigkeiten, von Neumann I. (1927-35) / Popper I. (1934-1953)

Der Lottofee, Versicherungsmathematikern und Naturwissenschaftlern war schon lange klar, dass eine enge Beziehung zwischen relativen Häufigkeiten und Wahrscheinlichkeiten besteht. Die Schule der *frequency dogmatists* postuliert den engsten Zusammenhang von allen: Identität. Demnach kann man etwa die Wahrscheinlichkeit, eine Vier zu würfeln, mit der Häufigkeit von Vierer-Würfen in einer *adäquaten* Abfolge von Würfen, dividiert durch die Gesamtzahl an Würfen mit dem Würfel, identifizieren. In ihrer einfachsten Form geht die Häufigkeits-Interpretation von einer *endlichen* Referenzklasse aus und auf Venn (1876) zurück, der in seiner Diskussion über die Geschlechterverteilung von Neugeborenen schlussfolgert: „probability is nothing but that proportion" (ebd. S. 84).[33] Dass diese operative Wahrscheinlichkeitsdefinition problematisch ist, liegt auf der Hand – man bedenke etwa das *Single-case-Problem*, um bloß eine unüberwindbare Barriere zu benennen.[34]

---

[32] Popper zieht dem wieder einen bedingten Wahrscheinlichkeitsausdruck vor.

Einige Frequentisten (d.h. Befürworter einer Häufigkeits-Interpretation) – in erster Linie Venn (1876), Reichenbach (1949) und R. von Mises (1957) – sind u.a. aus dem Grund dazu übergegangen, *unendliche* Referenzklassen zu erwägen, indem die Wahrscheinlichkeit p (A) eines Ereignisses A dem Grenzwert der relativen Häufigkeiten entsprechen soll.[35]

Verdeutlichen wir uns die beiden Positionen der Frequentisten. Nehmen wir die Wahrscheinlichkeit p (X) (X = A ∧ B) an, die als relative Häufigkeit interpretiert werden soll. Das bedeutet:[36]

1) Es existiert ein Ensemble $\mathcal{E}$, bestehend aus (endlich oder unendlich oder hypothetisch unendlich vielen) Ereignissen $N$, derart, dass

2) man für jedes Ereignis X eindeutig entscheiden kann und

3) ohne $\mathcal{E}$ dabei zu verändern, ob X der Fall ist oder nicht;

4) im Falle einer finiten Referenzklasse: $p(X) = \frac{\#(X)}{N}$ wobei $\#(X)$ der Zahl der Ereignisse in $\mathcal{E}$ entspricht, für welche X der Fall ist (und für $0 < N < \infty$);

4') im Falle einer infiniten Referenzklasse: $p(X) = \lim_{N \to \infty} \frac{\#(X)}{N}$.

Diese Annahmen kennzeichnen die Häufigkeits-Interpretationen der Wahrscheinlichkeit. Wie schneiden diese vor dem Hintergrund unserer Kriterien der Adäquatheit für Wahrscheinlichkeitsbegriffe ab? Gehen wir dieser Frage im Folgenden nach und legen zuerst ein besonderes Augenmerk darauf, inwieweit dem *Admissibility*-Kriterium bzw. den Kolmogorovschen Axiomen Rechnung getragen wird. Damit ist die Zeit für unsere Fallunterscheidung eingeläutet.

---

[33] Die frappierende (strukturelle) Ähnlichkeit zu dem frühen Wahrscheinlichkeitsbegriff der großen Mathematiker des 18.Jh. – darunter die Bernoulli-Familie, de Moivre, Bayes, Laplace – ist nicht zu leugnen; wohingegen allerdings die korrespondierende Interpretation dieser Wahrscheinlichkeit alle möglichen Ergebnisse eines Zufallsexperimentes zählt, berücksichtigt die finite Häufigkeits-Interpretation bloß tatsächliche Ergebnisse.

[34] Wenn nur einmal gewürfelt wird, dann ergibt sich eine relative Häufigkeit dafür, eine 4 zu würfeln, von 1 oder 0.

[35] Formal lässt sich diese Definition, die ihren Ursprung bei von Mises (1957) hat (vgl. die sogenannte ‚Limes-Definition'), so ausdrücken: $P(A) = \lim_{n \to \infty} r_n(A)$.

[36] gemäß von Mises (1957).

Prinzipiell ist voranzustellen, dass die Frequentisten die Axiome der Wahrscheinlichkeit als *empirische* Wahrheiten über relative Häufigkeiten von Ereignissen in der Welt ansehen. Insofern müssen diese Axiome nicht in jeder möglichen Welt Gültigkeit besitzen, damit die objektiven Wahrscheinlichkeitsbegriffe der Frequentisten überhaupt Früchte tragen können.

> […] the frequencist can adjust his/her views to accommodate new experimental evidence. If the laws of probability are empirical, not a-priori, they can be modified. The frequencist approach is thus less hard hit by quantum theory […]. (Pitowsky 1989, S. 186)

Das erlaubt es uns, unsere Fälle gewinnbringend (weil Transparenz schaffend) zu unterscheiden.

### 4.1.1. Fall 1:

Bei Geltung von Kolmogorovs Axiomen lässt sich konstatieren: Die finite Version stellt hinsichtlich der ersten beiden Grundregeln und auch bzgl. der finiten Additivität (siehe K3) zufrieden. In einer endlichen Referenzklasse können lediglich endlich viele Ereignisse geschehen, sodass nur endlich viele Ereignisse eine positive relative Häufigkeit haben können. Dennoch wird hier in Anspielung auf das *Single-case-Problem*, zu dem es zahlreiche verwandte Probleme gibt, behauptet, dass die finite Version der Häufigkeits-Interpretation aufzugeben sei, wofür zahlreiche Verweise in die Literatur als Beleg dienen können (vgl. etwa resümierend Hájek 2007b). Wie steht es um die Abwandlung der Interpretation für *unendliche* Referenzklassen?

### 4.1.2. Fall 2:

Die infinite Version bleibt vor Kritik nicht gefeit, sie stellt keine *admissible* Interpretation von Kolmogorovs Axiomatisierung dar. Sie verletzen das Additionsaxiom K3' (de Finetti 1972, §5.22).[37] Und weiter: „Indeed, the domain of definition of limiting relative frequency is not even a field, let alone a sigma field [a non-empty collection *F* of subsets of Σ is called a *sigma field*, C.H.]" (ebd. §5.8). Darum ist die infinite Version dem Fall 2, d.h. möglichen Welten zuzurechnen, in denen Nicht-Standard-Axiome der Wahrscheinlichkeit erfüllt sind, wofür die infinite Version der Häufigkeits-Interpretation *zulässig* sein kann.

Ad 2, *Ascertainability*-Kriterium: Dieses Kriterium, nach dem Wahrscheinlichkeiten zumindest grundsätzlich mithilfe einer bestimmten Methode ermittelbar sein müssen, bereitet der infiniten Version ebenfalls Schwierigkeiten. Bräuchte es nicht eine unendlich große Zahl an Würfen eines Würfels, um die genaue Wahrscheinlichkeit anzugeben, mit der mit einem bestimmten Würfel eine Vier gewürfelt wird? Und nicht nur für dieses Exempel – man denke z.B. an die sukzessive Abnutzung des Würfels, an Zeit- und Budgetrestriktionen des Spielers usw. – scheinen unendliche Versuchsserien, unendlich große Stichproben in der aktualen Welt nicht die Regel zu sein oder gar, nicht erstellt werden zu können. Jenseits des Empirismus könnte man versucht sein, Wahrscheinlichkeit mit einem hypothetischen oder kontrafaktischen Limes der relativen Häufigkeiten zu identifizieren; in dem Sinne, dass man sich hypothetische infinite Fortsetzungen der tatsächlichen Abfolge vorstellt. Doch auch wie lose und ungebunden mit der Wendung „in principle" in Salmons 2. Kriterium herumgespielt wird, das kann nicht darüber hinwegtäuschen, dass der Grenzwert im mathematischen Sinne gar nicht existiert.

---

[37] Dass sie gegen das finite Additionsaxiom K3 verstoßen, ist ohnehin offensichtlich, weil der Ergebnisraum bei infiniten Referenzklassen eben nicht mehr endlich ist, für welche Fälle die Erweiterung K3' eingeführt wurde.

Aber angenommen unsere Folge (z.B. der Würfe des Würfels) konvergiert – wofür Beschränktheit der Folge eine notwendige Voraussetzung ist –, so sind die Schwierigkeiten nicht behoben. Nicht nur dass jede Beobachtungsreihe irgendwann abgebrochen werden muss, ohne dass man sicher sein könnte, dass die relativen Häufigkeiten nah genug bei dem unbekannten Grenzwert liegen; sondern zudem würde man im Allgemeinen auch verschiedene ‚Wahrscheinlichkeiten' für das gleiche Ereignis A erhalten. Man stelle sich bspw. nachstehende unendliche Folge an Ergebnissen eines Würfelspiels vor: Augenzahl 4, 6, 3, 4, 4, 1, 2, 5, … Notieren wir uns der Klarheit willen die zugehörigen relativen Häufigkeiten dafür, eine Vier zu würfeln (= Ereignis A), nach jedem der gelisteten Würfe: 1/1, 1/2, 1/3, 2/4, 3/5, 3/6, 3/7, 3/8, …; diese Folge konvergiert *in the long run* gegen 1/6. Durch eine geeignete Neuanordnung der Ergebnisse können wir die Folge allerdings genauso gut gegen jeden anderen Wert im Intervall $[0, 1]$ konvergieren lassen.[38] [39]

Des Weiteren ist einzuwenden, dass relative Häufigkeiten nicht bloß von unterschiedlich auffassbaren Sequenzen innerhalb einer zu bestimmenden Referenzklasse abhängen, sondern somit auch von der Referenzklasse selbst, worin ebenfalls ein Problem für die infinite Version der Häufigkeits-Interpretation (aber nicht nur für diese, vgl. Hájek 2007a) liegt. Machen wir uns das bewusst, indem wir etwa die Wahrscheinlichkeit dafür bestimmen wollen, dass Ex-

---

[38] Vgl. dazu ausführlicher Hájek (2009 und 2007b), der dies für den analogen Fall der relativen Häufigkeit von geraden Zahlen unter den positiven ganzen Zahlen durchspielt. Wie er dort zeigt können sich die Frequentisten auch nicht dadurch retten, dass sie behaupten, *natürliche* Ordnungen (z.B. die zeitliche Ordnung) seien seinem künstlichem Arrangement strikt vorzuziehen.

[39] Historische Anmerkung: Auf vermeintliche Tücken der schwachen Formulierung des Gesetzes der großen Zahl bzw. der inversen Anwendung desselben machte bereits Leibniz in seiner Briefkorrespondenz mit Jakob Bernoulli aufmerksam. Vgl. dazu Leibniz (1988) sowie meine Arbeit über den Hauptsatz in der *Kunst des Vermutens*.

Bundeskanzler Frau Merkel ihr 80. Lebensjahr überschreitet.[40] Frau Merkel gehört der Klasse der Frauen, der Klasse der Kanzlerinnen der Bundesrepublik Deutschland, der Klasse der Nicht-Raucher, der Klasse der deutschen Physiker usw. an. Mutmaßlich variiert die relative Häufigkeit der Ü80 zwischen all diesen Referenzklassen. Welchen Wert beträgt dann aber die Wahrscheinlichkeit, dass Frau Merkel älter als 80 wird? Darauf scheint es keine eindeutige Antwort der Frequentisten zu geben, stattdessen kommen Wahrscheinlichkeiten-qua-Frau, -qua-Bundeskanzlerin-Nichtraucher usw. infrage.

Im Allgemeinen arbeiteten Reichenbach (1949) und von Mises (1957) an einer Lösung zu diesem Referenzklassen-Problem, indem sie unsere Aufmerksamkeit nur auf gewisse Folgen innerhalb gewisser Referenzklassen (mit gewissen wünschenswerten Eigenschaften) lenken möchten,[41] was einer kritischen Analyse aber nicht unbedingt standhält (vgl. Hájek 2009 und 2007b). Im Besonderen würde von Mises wohl unser Bsp. mit Fr. Merkel als ein zum Scheitern verurteilten Ermittlungsversuch einer *Single-case*-Wahrscheinlichkeit abtun, die in seinen Augen unsinnig seien: „We can say nothing about the probability of death of an individual even if we know his condition of life and health in detail. The phrase 'probability of death', when it refers to a single person, has no meaning at all for us" (ebd. S. 11). Zugespitzt formuliert gibt uns von Mises aber statt einer Lösung des Problems damit lediglich den Appell, das aufgemachte Problem zu ignorieren.[42]

Summa summarum: Die infinite Version der Häufigkeits-Interpretation wird den Anforderungen des *Ascertainability*- Kriteriums kaum gerecht, Wahrscheinlichkeitswerte bleiben unklar.

Ad 3, *Applicability*-Kriterium:

Anwendbarkeit auf Häufigkeiten: Entgegen naheliegender Erwartungen enttäuscht die infinite Version der Häufigkeits-Interpretation etwas im Hinblick auf dieses Subkriterium. Sie bindet die Wahrscheinlichkeit zu stark an den Grenzwert der relativen Häufigkeiten und vernachlässigt den Bezug zu endlichen Häufigkeiten: selbst bei unendlichen Folgen können die beiden auseinanderfallen – ein Laplacescher Würfel kann nach jedem Wurf und für alle Zeit mit der Augenzahl 4 oben liegen, was freilich sehr unwahrscheinlich ist (vgl. Hájek 2011).

---

[40] Bei diesem Bsp., das einmal nicht in das geschützte Biotop der Glücksspiele gebettet ist, leuchten ‚Wiederholungen' des Ereignisses, dass Frau Merkel älter als 80 wird, anders als bei einem wiederholten Wurf eines Würfels, gleich viel weniger ein.

[41] Vgl. weiter von Mises' (1957) Ausführungen zu seinem *Axiom of Convergence* und *Axiom of Randomness*.

[42] Man beachte, dass von Mises hier seinen eigenen Standpunkt unterminiert: der Ausdruck „probability of death" habe ebenso keine Bedeutung, falls er sich auf eine Milliarde Menschen (oder jegliche andere endl. Zahl) bezieht.

<u>Anwendbarkeit auf die Wissenschaften/die Quantenmechanik:</u> Der Siegeszug der Häufigkeits-Interpretationen in die Statistik ist wohlbekannt, aber welche Rolle spielt die infinite Version innerhalb der Quantenmechanik? Machen wir das wiederum an der Ansicht eines ihrer Protagonisten dazu fest: Von Neumann begrüßte in den Jahren 1927-1935 die Häufigkeits-Interpretation der Wahrscheinlichkeit. „[…] the Birkhoff and von Neumann concept of quantum logic could restore the harmonious classical picture: random events can be identified with the propositions stating that the event happens, and probabilities can be viewed as relative frequencies of the occurrences of the events" (Rédei 2009, S. 14).

Aber diese erneuerte Harmonie ist illusorisch. Die Annahmen 2 und 3 der vier die Häufigkeits-Interpretationen kennzeichnenden Bedingungen (siehe oben) können in der Quantenmechanik nicht aufrechterhalten werden. Annahme 3 muss man fallen lassen, wenn „entscheiden" (in 2) „messen" bedeutet, weil der Messvorgang / das Messgerät in Wechselwirkung mit dem zu messenden System steht, ergo auch das Ensemble Ɛ beeinflusst. Folglich, „[…] there is no single, fixed, well-defined ensemble in which to compute as relative frequencies the probabilities of *all* projections representing quantum attributes" (ebd.). Von Neumann war sich der Tragweite dieses Problems vollauf bewusst, glaubte aber (in 1932) eine *Ensemble*-Interpretation der Quantenwahrscheinlichkeit beibehalten zu können.[43]

Gestehen wir ihm einmal zu, dass eine *Ensemble*-Interpretation bedeutsam bliebe, wenn Annahme 3 – wie durch von Neumann vorgeschlagen – abgeschwächt werde; so ist er aber doch nach wie vor mit dem Problem konfrontiert, dass Annahme 2 in der Quantentheorie überhaupt keinen Sinn ergibt, sofern man die Position bezieht, dass (i) A ∧ B das gemeinsame Eintreten der (Quanten-) Ereignisse *A* und *B* repräsentiert (man bedenke X = A ∧ B, siehe oben) und (ii) deren gemeinsames Geschehen nicht durch einen Messprozess festgestellt werden kann (unabhän-gig vom Ensemble), wenn A und B nicht simultan messbar sind (Rédei 2009, S. 16).

Schließlich kann es darum, solange an der Häufigkeits-Interpretation der Wahrscheinlichkeit festgehalten wird, keine ,geeigneten nicht-kommutativen' Wahrscheinlichkeitsräume geben, die unabdingbar für die Quantentheorie sind – das ist schon aus Heisenbergs Unschärferelation ersichtlich. Von Neumann (1961) folgert daraus 1937 selbst: „This view, the so-called 'frequency theory of probability' has been very brilliantly upheld and expounded by R. von Mises. This view, however, is not acceptable to us […]." Und nachdem wir ansatzweise gesehen haben, dass auch Salmons drittes Kriterium insgesamt nicht ausreichend durch die Häufigkeits-Interpretation gewürdigt wird, dürfen wir hinzufügen: It is not acceptable to us either.

---

[43] Vgl. von Neumann (1932), insbesondere S. 300.

## 4.2. Propensität, Popper II. (1957)

Auch Popper gab die von ihm sehr geschätzte Häufigkeits-Interpretation 1953 (wie er selbst in 1959, S. 27 schreibt) auf, weil er sie – was nachvollzogen werden kann – als unbefriedigend in Verbindung mit dem Problem der Deutung der Quantentheorie empfand.[44] Hauptsächlich die Erkenntnisse aus den Doppelspaltexperimenten führten Popper zu einer neuen Interpretation der Wahrscheinlichkeit: Wahrscheinlichkeiten müssen nach seiner Überzeugung ‚physikalisch real' sein – „they must be physical propensities, abstract relational properties of the physical situation, like Newtonian forces, and 'real', not only in the sense that they could influence the experimental results, but also in the sense that they could, under certain circumstances (coherence), interfere, i.e. interact, with one another" (ebd. S. 28).

Die besondere Attraktivität seiner Theorie macht Popper darin aus, dass sie der Zuschreibung von *Single-Case*-Wahrscheinlichkeiten einen Sinn abgewinnen kann – wie es z.B. in Fällen der Form „die Wahrscheinlichkeit, dass dieses Atom des Elements Radium in 1600 Jahren zerfällt, beträgt 1/2" wünschenswert ist. In der Tat legt Popper (1957) seine Theorie der Propensität als eine Darstellung solcher quantenmechanischer Wahrscheinlichkeiten vor.[45]

In (1959) entwickelt er seine Theorie weiter: Die Wahrscheinlichkeit $p$ eines Ergebnisses von einem gewissen Typ ist eine physikalische Propensität, oder Disposition, oder Tendenz einer wiederholbaren Versuchsanordnung, Ergebnisse diesen Typs mit dem Grenzwert der relativen Häufigkeiten $p$ zu produzieren: „every experimental arrangement (and therefore every state of a system) generates physical propensities which can be tested by frequencies" (ebd. S. 38).

Wenn wir bspw. sagen, dass ein Würfel die Wahrscheinlichkeit 1/6 besitzt, mit der die Augenzahl 4 nach einem Wurf oben liegt, so meinen wir nach Popper, dass wir eine rekonstruierbare Versuchsanordnung haben (oder haben könnten), welche eine Tendenz besitzt, eine Serie von Ergebnissen zu erzeugen, in welcher der Grenzwert der relativen Häufigkeit von der Augenzahl 4 1/6 ist.

---

[44] Daneben erwähnt er das Single-Case-Problem als Motivation (vgl. ebd.).
[45] Dies bekräftigt Popper wieder in (1959) und wendet sich damit eindeutig gegen die Subjektivisten: „The main argument in favour of the propensity interpretation is to be found in its power to eliminate from quantum theory certain disturbing elements of an irrational and subjectivist character – elements which, I believe, are more 'metaphysical' than propensities and, moreover, 'metaphysical' in the bad sense of the word" (ebd. S. 31).

Die starke Abhängigkeit von oder die Rückführung auf Grenzwerte relativer Häufigkeiten birgt die Gefahr für Poppers Propensitätstheorie, von einem scheinbar überholten Von-Mises-Frequentismus hinunter gezogen zu werden.[46] Giere (1973) hingegen lässt explizit *Single-Case*-Propensitäten zu, ohne auf Häufigkeiten überhaupt einzugehen: Wahrscheinlichkeit sei lediglich eine Propensität einer rekonstruierbaren Versuchsanordnung, die eine Folge von Ergebnissen generiert. In dem Fall taucht jedoch das Gegenproblem zum Popperschen auf: Wie kriegen wir jetzt die gewünschte Verknüpfung zwischen Wahrscheinlichkeit und Häufigkeit (Hájek 2001)?

Es ist hilf- und aufschlussreich Gillies (2000) darin zu folgen, zwischen Theorien der *Long-run*-Propensitäten und der *Single-Case*-Propensitäten zu differenzieren:

> A long-run propensity theory is one in which propensities are associated with repeatable conditions, and are regarded as propensities to produce in a long series of repetitions of these conditions frequencies which are approximately equal to the probabilities. A single-case propensity theory is one in which propensities are regarded as propensities to produce a particular result on a specific occasion.[47] (ebd. S. 822)

Hacking (1965) und Gillies (ebd.) offerieren *Long-run*-Theorien der Propensität (wohlbemerkt für lediglich endliche *long runs*), Fetzer (1982) und Miller (1994) plädieren für die *Single-Case*-Varianten. Unabhängig davon, welcher Position man den Vorzug geben mag, bietet sich wie bei der Diskussion der Häufigkeits-Interpretationen unsere Fallunterscheidung an, bevor wir die Theorien der Propensität mithilfe unserer Kriterien evaluieren möchten. Auch hier kann die Gegenüberstellung von möglichen Welten, in denen Kolmogorovs Axiome erfüllt respektive unerfüllt sind, einen Mehrwert stiften, weil Propensitäts-Interpretationen (genau wie die Häufigkeits-Interpretationen) nicht voraussetzen, dass diese Axiome notwendige Wahrheiten sind.

---

[46] Popper (1959) überspielt diesen Vorwurf leichtfertig, indem er schreibt (S. 26): „[…] I believe that it is possible to construct a frequency theory of probability that avoids all the objections which have been raised and dis­cussed." Diese Entgegnung bleibt schwach, weil völlig unklar ist, wie ein solcher Frequentismus aussehen könnte.

[47] Man nehme zur Kenntnis, dass Propensitäten kategorisch verschiedene Dinge sind, je nach dem, welche Theorie man favorisiert. Gemäß den *Long-run*-Theorien kommen Propensitäten Tendenzen gleich, relative Häufigkeiten mit bestimmten Werten hervorzubringen, wobei die Propensitäten allerdings nicht die Wahrscheinlichkeitswerte selbst sind. Nach den *Single-Case*-Theorien entsprechen die Propensitäten den Wahrscheinlichkeitswerten.

### 4.2.1. Fall 1:

Für mögliche Welten, wo die Standard-Axiome der Wahrscheinlichkeit gelten, können wir festhalten: Auf der einen Seite ist *prima facie* zumindest unklar, ob *Single-Case*-Theorien dem Wahrscheinlichkeitskalkül gehorchen oder nicht, d.h. es ist schleierhaft, ob sie in Bezug auf Kolmogorovs Axiomensystem *admissible* sind oder nicht. Angesichts dieser unbefriedigenden Ambivalenz könnte man nun auch einfach vorschreiben, dass sie Salmons erstem Kriterium genügen; vielleicht indem man diese Festsetzung als Teil der impliziten Definition von Propensitäten mit aufnimmt – bloß aber festzusetzen, was ein fliegendes Spaghettimonster ist, reicht nicht aus, um zu zeigen, dass ein fliegendes Spaghettimonster existiert. Die dann aufkommende Frage ist also, ob solche postulierten Propensitäten, sodass die *Single-Case*-Propensitäts-Interpretationen relativ zu Kolmogorovs Axiomen zulässig sind, existieren oder nicht. In der Tat weisen Kritiker darauf hin, dass wir zu wenig darüber wissen, was Propensitäten sind, um derartige Fragestellungen beurteilen zu können (vgl. Hitchcock 2002).

Humphreys (1985) präsentiert ein richtungsgebendes Argument – bekannt geworden als Humphreys' Paradoxon –, nach dem (*Single-Case*-) Propensitäten nicht Kolmogorovs Wahrscheinlichkeitskalkül gehorchen. Die Idee ist, dass der Wahrscheinlichkeitskalkül den Satz von Bayes impliziert, der es uns gestattet, eine konditionale Wahrscheinlichkeit zu ‚invertieren‘:[48] [49]

$$p(A|B) = \frac{p(B|A) \times p(A)}{p(B)}$$

Propensität scheine dagegen ein Maß für ‚kausale Tendenz‘ zu sein und da die Kausalrelation asymmetrisch ist – d.h. die Kausalität hat eine feste Richtung, auf eine Ursache folgt ihre Wirkung (kausal, nicht zeitlich)[50] und nicht umgekehrt –, seien diese Propensitäten, die eben nach den *Single-Case*-Theorien gleich den Wahrscheinlichkeiten sind, angeblich nicht zu invertieren. Gegeben etwa ein rapides Absinken des Luftdrucks mag eine (nicht-triviale) Propensität haben, dass ein Gewitter aufzieht; wenngleich es scheinbar keinen Sinn macht zu sagen, ein gegebenes aufziehendes Gewitter hat eine (nicht-triviale) Propensität, dass der Luftdruck rapide abgesunken ist. Also haben wir ein Argument zur Hand, dass Propensitäten – was auch immer sie seien mögen – nicht den geltenden Axiomen gerecht werden dürfen.

---

[48] Angenommen es gibt zwei jeweils mit grünen und roten Kugeln gefüllte Urnen A und B. Das feste Verhältnis zwischen den roten und grünen Kugeln ist uns für beide Urnen bekannt. Eine Urne wird zufällig ausgewählt. Nehmen wir weiter an, es gibt ein Ereignis R: Ziehen einer roten Kugel aus einer Urne. Die Wahrscheinlichkeit, eine rote Kugel zu ziehen, beträgt für Urne A 0,6 und 0,4 für Urne B. Somit ergeben sich folgende konditionale Wahrscheinlichkeiten: p (R/A) = 0,6 und p (R/B) = 0,4. Stellen wir uns aber die Frage, was die Wahrscheinlichkeit ist, dass die gewählte Urne A bzw. B ist, *gegeben* dass eine rote Kugel gezogen wurde. Damit fragen wir nach den invertierten konditionalen Wahrscheinlichkeiten p (A/R) bzw. p (B/R), worauf uns der Satz von Bayes eine gute Antwort gibt.

[49] Dies ist nur eine mögliche Schreibweise des Satzes von Bayes. Die auf S. 9 gegebene Definition der konditionalen Wahrscheinlichkeit (siehe Carnaps *confirmation function* c (–, –)) ist äquivalent dazu.

[50] Die kausale Asymmetrie sollte nicht mit der zeitlichen Asymmetrie verwechselt werden, auch wenn in der Literatur sowohl ersteres auf letzteres (vgl. Hume 1999, Suppes 1970) als auch letzteres auf ersteres (vgl. Reichenbachs *kausale Theorie der Zeit*, in Reichenbach 1928) gegründet wurde. Vgl. auch meine Arbeit über rück-wirkende Kausalität und ihre Relevanz für Newcombs Problem.

Dieses Argument ermunterte Fetzer und Nute (in Fetzer 1981) dazu, einen alternativen ‚probabilistic causal calculus' einzuführen, der sich vom Kolmogorovschen Kalkül recht deutlich abhebt. Auch wenn einige Autoren Humphreys' Schluss nicht zustimmen und etwa argumentieren, dass sein Paradox uns keineswegs nicht-Kolmogorovsche Propensitäten aufzwingt (vgl. z.B. Gillies 2000), können wir diese Debatte (v.a. aus Platzgründen) nicht beleuchten und nehmen Humphreys' Argument ohne nähere Prüfung als korrekt an. Somit stellen *Single-Case*-Theorien der Propensität keine *admissible* Interpretation von Kolmogorovs Axiomatisierung dar, was mit einbezieht, dass zumindest die *Single-Case*-Varianten unter Fall 2 zu listen sind.

Auf der anderen Seite scheinen solche Theorien, die Propensitäten an Häufigkeiten binden (wie Popper es tut), ohnehin keine *admissible* Interpretation des (vollen) Wahrscheinlichkeitskalküls zu sein und das klarerweise aus den gleichen Gründen, die wir schon zuvor bei der Besprechung der Häufigkeits-Interpretationen der Wahrscheinlichkeit herausgearbeitet haben.

### 4.2.2. Fall 2:

Wir können also zusammenfassen, dass sowohl *Long-run*-Theorien als auch *Single-Case*-Theorien der Propensität in Welten, die durch Kolmogorovs Axiomatisierung gekennzeichnet sind, in Fall 1 also, nicht zulässig sind und für Welten, wo andere Axiome für Wahrscheinlichkeitsverteilungen gelten, – diese gehören zu Fall 2 – zulässig sein können – z.B. ist die *Single-Case*-Theorie von Fetzer und Nute (in Fetzer 1981) für die von ihnen entwickelte Axiomatisierung zulässig. Wie überzeugend sind aber Propensitäts-Interpretationen gemäß unseren Kriterien [2] und [3] zur Sicherung der Adäquatheit von Wahrscheinlichkeitsbegriffen?

Ad 2, *Ascertainability*-Kriterium: Der Forderung nach der prinzipiellen Bestimmbarkeit von Wahrscheinlichkeiten mithilfe einer geeigneten Methode kommen die *Long-run*-Theorien nicht hinlänglich nach, was aus Abschnitt 4.1.2. zu Genüge hervorgeht.

Die *Single-Case*-Theorien ereilt kein besseres Urteil. Laut diesen Interpretationen entspricht eine Wahrscheinlichkeit einer Propensität (einer rekonstruierbaren Versuchsanordnung, die eine Folge von Ergebnissen produziert); so weit, so gut. Darauf lässt sich mindestens zweierlei erwidern: Zu behaupten, dass eine experimentelle Anordnung eine bestimmte Tendenz besitzt, setzt erstens eine gewisse Stabilität oder Gleichförmigkeit in der Funktionsweise dieses Setups voraus, die man als eine Form der *Konformitätsthese* („die Natur sei uniform") lesen kann, wovon schon Hume (1999) sagte, dass sie weder *a priori* noch *a posteriori* gewusst werden könne.[51]

Zweitens wissen wir eben nicht genug über Propensitäten, um entsprechende Wahrscheinlichkeiten ermitteln zu können. Angenommen ein Würfel habe eine schwache Propensität, mit der Augenzahl 4 oben zu landen – sagen wir, der Wert 1/6 messe diese Propensität; doch Hitchcock (2002) betont: „calling this property a 'propensity' of a certain strength does little to indicate just what this property is" und noch weniger, wie sie gemessen werden kann.

Zugespitzt formuliert sind Propensitäts-Interpretationen (für den long run *und* den single case) der gravierenden Schwäche bezichtigt, leere Darstellungen der Wahrscheinlichkeit zu geben, à la Molières ‚dormative virtue' (Sober 2000, S. 64).

Ad 3, *Applicability*-Kriterium:
<u>Anwendbarkeit auf die Wissenschaften/die Quantenmechanik:</u> Aufgrund dieser berechtigten Anklage haben die Propensitäts-Theorien auch erhebliche Schwierigkeiten, Anwendung und Eingang in die Wissenschaft zu finden. Gillies (2000) wirft bspw. *Single-Case*-Propensitäten vor, dass Aussagen über sie nicht überprüfbar und dass sie „metaphysical rather than scientific" seien (ebd. S. 825). Dieser Vorwurf strahlt auch in Richtung der *Long-run*-Propensitäten aus, die vermutlich von anderer Art als die überprüfbaren relativen Häufigkeiten sind. Jedenfalls artikuliert sich auch hier, dass „propensity accounts have been criticized for being unacceptably vague" (Hájek 2001, S. 372).

Alles in allem ist darum nicht zu sehen, wie Propensitäts-Interpretationen und die Quantentheorie nachhaltig füreinander fruchtbar gemacht werden können, wofür der Mangel an belastbaren Literaturbeiträgen, die sich für Propensitäten als den adäquaten Wahrscheinlichkeitsbegriff einsetzen, als Evidenz dienen mag. Auch hinsichtlich der klassischen Wahrscheinlichkeitstheorie glänzen die Theorien von Popper & Co. wenig, was ihre schlechte Bilanz gemäß unseren Kriterien widerspiegelt.

---

[51] Die Berufung auf Grenzwerttheoreme / Gesetze der großen Zahlen hilft hier nicht weiter. Vgl. Hájek (2007b).

So lautet unser abschließendes Urteil über Propensitäten vor dem Hintergrund unserer *Kernfrage*, für das kein weiteres stützendes Material gesammelt werden soll.[52] Stattdessen wollen wir lieber ein Gesamtfazit ziehen.

## 5. Bewertung, Ausblick und abschließende Bemerkungen

Die vorstehende Analyse mit einer Darstellung bzw. dem bloßen Ansprechen von Positionen und einhergehenden Problemen, mit den häufigen Verweisen in die Literatur unterstreicht, dass noch viel Arbeit in die Suche nach einem akzeptablen Wahrscheinlichkeitsbegriff investiert werden sollte. Vorläufig bleibt die Antwort auf unsere *Kernfrage* eine negative: Keine der vorgestellten Interpretationen der Wahrscheinlichkeit, die von ihren jeweiligen Anhängern meist als die führende zelebriert werden, stellt einen kritischen Leser hinsichtlich der klassischen Wahrscheinlichkeitstheorie oder der Quantentheorie zufrieden.

Weder subjektive Wahrscheinlichkeitsbegriffe noch objektive hielten der präzisen, da durch Fallunterscheidungen geprägten, Analyse stand, wobei die Folgen unserer Erkenntnisse besonders verheerend für das Carnapsche Lager waren. Jede Interpretation, welche untersucht wurde, scheint zwar einen oder mehrere wesentliche Aspekte der Wahrscheinlichkeit zu erfassen, jedoch enttäuschen sie alle in anderer wesentlicher Hinsicht, was durch ihre Performance entsprechend den in Abschnitt 2 lancierten Kriterien reflektiert wird. Vielleicht ergäbe sich ein kohärentes Gesamtbild, wenn man auf eine Art *Patchwork* insistieren würde. Demnach könnte man die obigen Interpretationen als komplementär ansehen, wiewohl es an allen Stellen Ausbesserungen und Verfeinerungen bedürfte, um den methodischen und konzeptionellen Schwierigkeiten zu begegnen. Sicherlich werden dabei aber sowohl objektive als auch subjektive Elemente der Wahrscheinlichkeit bewahrt.

Künftig sollte an zwei Fragestellungen respektive Stellschrauben angesetzt werden: Die erste betrifft die Axiomatisierung der Wahrscheinlichkeit: Gibt es aus der Schar bestehender (oder noch erscheinender) Axiomatisierungen eine ‚beste‘? Macht die vorherige Frage überhaupt

---

[52] Um Doppelungen zu vermeiden – im Zusammenhang mit Häufigkeits-Interpretationen wurde schon einiges genannt, was hier offensichtlich auch angeführt werden könnte (siehe oben) – und da die bereits aufgedeckten schwerwiegenden Schwachstellen wohl für die abschließende Bewertung der Propensitäten ausgereicht haben, sollen weitere Subkriterien zu Salmons *Applicability*-Kriterium ausgespart werden. Weiterführend äußert sich z.B. Lewis (1980) mit seinem ‚Principal Principle‘ zur Anwendbarkeit auf rationale Glaubensgrade.

Sinn? Was spricht für, was spricht gegen diese oder jene Axiomatisierung? Sollten wir verschiedene Axiomatisierungen der Wahrscheinlichkeit als gleichberechtigt zulassen? Etc.

Insofern wird es (für Mathematiker und auch für Philosophen) wichtig, konkurrierende, sich ergänzende oder für unterschiedliche Felder zutreffende Theorien der Wahrscheinlichkeit zu identifizieren, zu vergleichen, abzuwägen und zu evaluieren. Vor dem Hintergrund unserer *Kernfrage* wäre v.a. interessant, ob wir verschiedene Wahrscheinlichkeiten haben – etwa eine klassische Wahrscheinlichkeit (für z.B. den Bereich der Glücksspiele) und eine Quantenwahrscheinlichkeit – oder ob es eine und dieselbe Wahrscheinlichkeit ist; ob wir bspw. radikal Physikern wie Dirac, Wigner und Feynman folgen sollten oder könnten, die sogar *negative* Wahrscheinlichkeiten dulden, oder ob es in diesem Fall sinnfrei wäre, überhaupt von Wahrscheinlichkeiten zu sprechen.[53]

Mit der zweiten Stellschraube werden – wie in dieser Arbeit getan – die Interpretationen von diesem oder jenem Wahrscheinlichkeitskalkül in Betracht genommen. Neben der *Kernfrage* kann etwa zu bedenken gegeben werden: Gibt es nur einen einzigen (z.B. gemäß Salmons Kriterien) adäquaten Wahrscheinlichkeitsbegriff? Gibt es überhaupt einen? Gleicht er vielleicht einem *Patchwork*?[54]

Es zeichnet sich bereits eine Rehabilitation objektiver und subjektiver Wahrscheinlichkeitsbegriffe ab.[55] Dennoch bleibt die Interpretation klassischer Wahrscheinlichkeiten ein brisantes Unterfangen und wie Rédei (2009) gerne ergänzt sehen würde: „the interpretation of quantum probability remains a much debated issue even today" (ebd. S. 20).

---

[53] Bei Bejahung von letzterem würden wir an unserem intuitiven oder vortheoretischen Begriff der Wahrscheinlichkeit festhalten (zumindest in dem Sinne, dass Wahrscheinlichkeitswerte nicht negativ sein können) und eine Korrektur an der radikalen Wahrscheinlichkeitstheorie vornehmen. Die Parallele zu dem von Goodman (1983) stark gemachten Überlegungsgleichgewicht (bei dem Sozialphilosophen John Rawls, dem es um die Begründung der Moral geht, heißt es in der Tat *reflective equilibrium*) im Zusammenhang mit unserer induktiven Praxis ist gut erkennbar.

[54] Eine andere Facette dieser zweiten Fragestellung belangt den adäquaten Wahrscheinlichkeitsbegriff für die Quantentheorie an und damit auch die vielen Interpretationen derselben: Welche ist die angemessene Interpretation? Welcher Wahrscheinlichkeitsbegriff passt zu welcher Interpretation der Quantenmechanik? Usw.

[55] Vgl. weiterführend Pearl (2000), Maher (2001) und Schervish/Seidenfeld/Kadane (2000).

## 6. Literatur

## Über den Autor

Dr. Christian Hugo Hoffmann ist Geschäftsführer und Gründer von Hoffmann Economics, einer Investment- und Investmentberatungsboutique mit Schwerpunkt auf Venture Capital. Er hat an der Universität St. Gallen (Schweiz) in Management promoviert, arbeitete als Assistenzprofessor für Finance an der Universität Liechtenstein und schreibt derzeit seine zweite Doktorarbeit über die Ethik der Künstlichen Intelligenz. Neben seiner Leidenschaft für Künstliche Intelligenz (KI) im akademischen Bereich ist Christian im Herzen ein Tech-Unternehmer mit drei Software-Start-ups in Deutschland, der Schweiz und Malawi. Darüber hinaus war er stellvertretender Direktor und Leiter der KI-Abteilung des Swiss Fintech Innovation Lab in Zürich, Direktor von Startup Grind Genf und erfüllt weiterhin seine Rolle als Start-up-Investor, Coach und Mentor in verschiedenen Programmen (u.a. MassChallenge, Vroom, Kickstart Accelerator) und mit Beteiligungen an mehreren Tech-Start-ups. Weitere Informationen zu seiner Person unter: https://www.christian-hugo-hoffmann.com/.

**Name:** Dr. Christian Hugo Hoffmann
**Position:** Geschäftsführer & Gründer
**Zugehörigkeit:** Hoffmann Economics
**E-Mail:** christian@hoffmann-economics.com
**ORCID:** 0000-0002-9822-5034